\documentstyle[12pt]{article}
\begin{document}
\noindent
\begin{center}
{\Large {\bf Hadamard States and \\Two-dimensional Gravity\\}}

\vspace{2cm}

${\bf H.~Salehi}$\footnote{e-mail:h-salehi@cc.sbu.ac.ir},
${\bf Y.~Bisabr}$\footnote{e-mail:y-bisabr@cc.sbu.ac.ir}\\

\vspace{0.5cm}
{\small {Department of Physics, Shahid Beheshti University, Evin,
Tehran 19839,  Iran.}}\\
\end{center}
\vspace{1cm}
\begin{abstract}
We have used a two-dimensional analog of the Hadamard state-condition
to study the
local constraints on the two-point function of a linear quantum field
conformally
coupled to a two-dimensional gravitational background.  We develop a dynamical
model in which the determination of the state of the quantum field is essentially
related to the determination of a conformal frame.  A particular conformal
frame
is then introduced in which a two-dimensional gravitational equation is
established.
\end{abstract}
\vspace{4cm}
\section{Introduction}
The subject of quantum field theory in curved spacetime studies the
interrelation between quantum theory and spacetime geometry in the
approximation that the geometry is remained as a classical background.  This
subject, however, contains some basic problems concerning inherent
ambiguities in the definition of physical states associated with a quantum
field.  In the absence of a gravitational
background, it is always possible
to use the Poincare symmetries to obtain a physical vacuum (the state of
lowest energy).  One may then assume that the physically admissible states
most naturally arise as local excitations of this state.  In the presence
of a curved spacetime, however, this procedure does not apply,
because on a general curved sapcetime one may not find a global
symmetry.  In this case the
problem concerning the determination of the local states
and the
role of the global features of spacetime is of obvious importance.\\
The Hadamard state-condition provides a framework in which we may improve
our understanding in this context.  In this framework one postulates that
the short-distance singularity structure of the two-point function of a
linear quantum field is represented by the Hadamard
ansatz \cite{h}.  In this prescription, however, there exist
problems in the specification of the state-dependent
part of the two-point function.  For characterizing the physical states, it
therefore seems to be essential to find out a suitable scheme for the treatment
of these problems.  In the present paper
we shall consider this issue for a quantum scalar field conformaly coupled
to a two-dimensional gravitational background.\\
The organization of this paper is as follows: In section 2, we use
the Hadamard ansatz for the
derivation of the local constraints imposed on the state-dependent part of the
two-point function.  In section 3, we develop a
dynamical model
in order to analyze these constraints.
This dynamical model, which is a two-dimensional analog of a model considered
in the previous publication \cite{sbg}, uses a conformal invariant
C-number field to characterize
the stress-tensor associated with the local states.  The
introduction of this scalar field into the
analysis leads to a basic connection between the specification of
the local states and the
determination of a conformal frame.  There is a general consistency
requirement on the behavior of the C-number field which necessitates the
use of a
conformal frame in which the trace of the stress-tensor is
measured by a cosmological constant.  This leads to the establishment of a
two-dimensional gravitational equation in this frame.  In section 4, we
offer some concluding remarks.  Throughout the following we shall
work in units in which $\hbar=c=1$ and
follow the sign conventions of Hawking and Ellis \cite{he}.
\section{Hadamard states in two spacetime dimensions}
We consider a free massless quantum scalar field $\phi(x)$ propagating in a
two-dimensional gravitational background with the action
functional (In the following the semicolon and $\nabla$ denote covariant
differentiation)
\begin{equation}
S[\phi]=-\frac{1}{2} \int d^{2}x~ {\sqrt {- g}} ~~g_{\alpha\beta}
\nabla^{\alpha}\phi \nabla^{\beta}\phi~.
\label{2.1}
\end{equation}
This action gives rise to the field equation
\begin{equation}
\Box \phi(x)=0~.
\label{2.2}
\end{equation}
A state of $\phi(x)$ is characterized by a hierarchy of wightman functions
\begin{equation}
\langle \phi(x_1),...,\phi(x_n) \rangle~.
\label{2.11}
\end{equation}
We are primarily interested in quasi-free states, i.e., states for which
the truncated
n-point functions vanish for $n>2$.  Such states may be characterized
by their two-point functions.  In
a linear theory the antisymmetric part of the two-point function
is common to all states in the same representation.  It is just
the universal commutator function.  Thus, in our case all the relevant
information about the state-dependent part of the two-point function are
encoded in its symmetric part, denoted in the following by $G^{+}(x,x')$,
which satisfies Eq.(2) in each argument.  Equivalence
principle suggests that the leading
singularity of $G^{+}(x,x')$ should have a close correspondence
to the singularity
structure of the two-point function of a free massless field in a
two-dimensional Minkowski
spacetime.  One may therefore assume that for $x$ sufficiently close
to $x'$ the function $G^{+}(x,x')$ can be written as
\begin{equation}
G^{+}(x,x')=-\frac{1}{4\pi}\ln \sigma(x,x')+F(x,x')~,
\label{2.3}
\end{equation}
where $\sigma(x,x')$ is one-half the square of the geodesic distance
between $x$ and $x'$, and $F(x,x')$
is a regular function.  This may be viewed as a two-dimensional analog of the
Hadamard ansatz\cite{h}.  It should be remarked that in general there
is a missing mass scale in the expression (\ref{2.3}) emerging from the
fact that the argument of the logarithm must be dimensionless.  We shall
return to this point in the next section.\\
The function $F(x,x')$ satisfies
a general constraint obtaining from the symmetry condition
of $G^{+}(x,x')$ and requiring that the expression (\ref{2.3}) satisfies
the wave equation (\ref{2.2}).  The study of this constraint has obviously
a particular significance for analyzing the state-dependent part of the
two-point function.  We therefore present its derivation as follows:
Applying Eq.(\ref{2.2}) on (\ref{2.3}) leads to
\begin{equation}
\sigma(x,x') \Box F(x,x')=\frac{1}{4\pi}(\sigma ^{;\alpha }_{~~;\alpha }-2)~,
\label{2.4}\end{equation}
in which we have used the definition of $\sigma (x,x')$, namely
\begin{equation}
\sigma (x,x') =\frac{1}{2}g_{\alpha \beta} \sigma ^{;\alpha } \sigma ^{;\beta}~.
\label{2.5}\end{equation}
The explicit form of the right hand side of (\ref{2.4}) can be specified
by writing the covariant expansion of $\sigma ^{;\alpha\beta}$ \cite{ch}
\begin{equation}
\sigma ^{;\alpha \beta}=g^{\alpha \beta}-\frac{1}{3}R^{\alpha ~~\beta~~}
_{~~\gamma~~\delta}\sigma ^{;\gamma } \sigma ^{;\delta }+\frac{1}{12} R^{\alpha~~\beta~~}
_{~~\gamma~~\delta;\rho }\sigma ^{;\gamma }\sigma ^{;\delta }
\sigma^{;\rho}+O(\sigma^2)~.
\label{2.6}\end{equation}
Taking the trace of this relation and using the fact that in a two-dimensional
spacetime the Einstein tensor identically vanishes
\begin{equation}
R_{\alpha \beta}=\frac{1}{2}g_{\alpha \beta}R~,
\label{2.7}\end{equation}
leads to
\begin{equation}
\sigma ^{;\alpha}_{~~;\alpha }=2-\frac{1}{6}g_{\gamma \delta }
R \sigma ^{;\gamma }\sigma ^{;\delta }+\frac{1}{24} g_{\gamma\delta} R_{;\rho }
\sigma ^{;\gamma }\sigma ^{;\delta } \sigma^{;\rho}+O(\sigma^2)~.
\label{2.8}\end{equation}
Combining (\ref{2.4}) and (\ref{2.8}), yields
\begin{equation}
\sigma(x,x') \Box F(x,x')=-\frac{1}{24\pi}g_{\gamma \delta }
R \sigma ^{;\gamma }\sigma ^{;\delta }+\frac{1}{96\pi} g_{\gamma\delta}
R_{;\rho }\sigma ^{;\gamma }\sigma ^{;\delta } \sigma^{;\rho}+O(\sigma^2)~.
\label{2.9}\end{equation}
Substituting (\ref{2.5}) and the covariant expansion
of $F(x,x')$ \cite{b}, namely
\begin{equation}
F(x,x')=F(x)-\frac{1}{2}F_{;\alpha}(x) \sigma^{;\alpha}+\frac{1}{2}
F_{\alpha \beta}(x) \sigma^{;\alpha} \sigma^{;\beta}
+\frac{1}{4}\{\frac{1}{6}F_{;\alpha\beta\gamma}(x)-
F_{\alpha\beta;\gamma}(x)\}\sigma^{;\alpha}\sigma^{;\beta}\sigma^{;\gamma}+
O(\sigma^{2})~,
\label{2.10}\end{equation}
into (\ref{2.9}), one finds
$$
\frac{1}{2}F^{\eta}_{~~\eta}(x)g_{\gamma \delta } \sigma^{;\gamma}
\sigma^{;\delta}
+\frac{1}{2}g_{\gamma \delta}[F_{\alpha \beta}^{~~;\alpha }(x)-\frac{1}{2}
F^{\eta}_{~~\eta;\beta}(x)+\frac{1}{12}(\Box F(x))_{;\beta}-\frac{1}{3}
\Box(F_{;\beta}(x))-\frac{1}{12}RF_{;\beta}(x)]\sigma^{;\gamma}
\sigma^{;\delta}\sigma^{;\beta}+O(\sigma ^2)
$$
\begin{equation}
~~~=-\frac{1}{24\pi}g_{\gamma \delta }
R \sigma ^{;\gamma }\sigma ^{;\delta }+\frac{1}{96\pi} g_{\gamma\delta}
R_{;\beta }
\sigma ^{;\gamma }\sigma ^{;\delta } \sigma^{;\beta}+O(\sigma^2)~.
\label{2.11}\end{equation}
We may now compare term by term up to the third order
of $\sigma^{;\alpha } $ to
obtain the desired relations
\begin{equation}
F^{\eta}_{~~\eta}(x)=-\frac{1}{12\pi}R~,
\label{2.12}\end{equation}
\begin{equation}
F_{\alpha \beta}^{~~;\alpha}(x)-\frac{1}{2}
F^{\eta}_{~~\eta;\beta}(x)+\frac{1}{12} (\Box F(x))_{;\beta}
-\frac{1}{3}\Box(F_{;\beta}(x))-\frac{1}{12}RF_{;\beta}(x)=
\frac{1}{48\pi}R_{;\beta}~.
\label{2.13}\end{equation}
The equations (\ref{2.12}) and (\ref{2.13}) are to be regarded as
two independent constraints imposed on the state-dependent
part of the two-point function.  We may combine them to obtain
\begin{equation}
F_{\alpha \beta}^{~~;\alpha}(x)+\frac{1}{12} (\Box F(x))_{;\beta}
-\frac{1}{3}\Box(F_{;\beta}(x))-\frac{1}{12}RF_{;\beta}(x)=
-\frac{1}{48\pi}R_{;\beta}~.
\label{2.14}\end{equation}
It should be noted that in the derivation
of this constraint we have
used the covariant expansion of $F(x,x')$ and $\sigma ^{;\alpha \beta}$ only
up to the second order
terms.  In general there exist additional constraints on the higher
order expansion terms.   However in our analysis we shall neglect these higher
order constraints and concentrate our attention on (\ref{2.14}).\\
\section{The large scale behavior of local states}
In the first step of analyzing the constraint (\ref{2.14}) we note that it
can be written as a total divergence.  In fact we may use the
relation (\ref{2.7}) and the differential identity
\begin{equation}
\Box(F_{;\beta}(x))=(\Box F(x))_{;\beta}+R_{\alpha\beta}F^{;\alpha}(x)~,
\end{equation}
to get the equation
\begin{equation}
\nabla^{\alpha} \Sigma_{\alpha\beta}=0~,
\label{3.2}\end{equation}
where
\begin{equation}
\Sigma_{\alpha\beta}=\Sigma_{\alpha\beta}^{(0)}
+\Sigma_{\alpha\beta}^{(1)}~,
\label{3.3}
\end{equation}
and
\begin{equation}
\Sigma_{\alpha\beta}^{(0)}=
\frac{1}{2}(F_{;\alpha\beta}(x)-\frac{1}{2}g_{\alpha\beta} \Box F(x))
-(F_{\alpha\beta}+\frac{1}{24\pi}g_{\alpha\beta}R)~,
\label{3.4}
\end{equation}
\begin{equation}
\Sigma_{\alpha\beta}^{(1)}=\frac{1}{48\pi}g_{\alpha \beta}R~.
\label{3.43}
\end{equation}
The tensor $\Sigma_{\alpha\beta}$ is decomposed into a traceless
part, which is denoted by $\Sigma_{\alpha\beta}^{(0)}$, and the
tensor $\Sigma_{\alpha\beta}^{(1)}$
which leads to the trace anomaly
in two-dimensions\cite{bd,cf}.  The principal
strategy now is to relate
the tensor $\Sigma_{\alpha\beta}^{(0)}$ to the traceless stress-tensor of a
conformal invariant scalar field which is given by \cite{bd}
\begin{equation}
T_{\alpha\beta}=\nabla_{\alpha}\psi \nabla_{\beta}\psi-\frac{1}{2}
g_{\alpha\beta} \nabla_{\gamma}\psi \nabla^{\gamma}\psi~,
\label{3.8}\end{equation}
where $\psi(x)$ is a C-number scalar field satisfying the dynamical equation
\begin{equation}
\Box \psi(x)=0~.
\label{3.6}
\end{equation}
In this case the tensor $\Sigma_{\alpha \beta}$ takes the form
\begin{equation}
\Sigma_{\alpha\beta}=T_{\alpha\beta}+\frac{1}{48\pi}g_{\alpha \beta}R~.
\label{3.7}
\end{equation}
This relation indicates
how the local states are related
to the background geometry.  Taking its trace we obtain
\begin{equation}
\Sigma^{\alpha }_{\alpha }=\frac{1}{24\pi}R~,
\label{3.9}
\end{equation}
which is the trace anomaly of a two-dimensional quantum stress-tensor of
a conformal invariant scalar field.  This raises the possibility of regarding
$\Sigma_{\alpha \beta}$ as the quantum stress-tensor induced by the two-point
function.  However by comparing (\ref{3.7}) with (\ref{3.2}) we observe
that the existence of the trace anomaly on the background metric is in conflict
with the interpretation of $T_{\alpha \beta}$ as a conserved
stress-tensor.  This observation reveals that the
connection between the local properties of the state-dependent part of the two-point
function with those of
a C-number scalar field, $\psi(x)$, can not be consistently established on
the background
metric.  This apparent discrepancy, however, may be removed by appealing to the
conformal symmetry of the above model.  The conformal invariance of
Eq.(\ref{3.6}) implies that
at dynamical level it is not possible to single out a particular
configuration for $\psi(x)$ among many different conformally related
configurations.  In order
to determine a configuration one has to choose a particular
conformal frame.  This therefore raises the question of which of these
conformal frames corresponds to the physical one.  In general the choice of a
conformal frame depends on the physical conditions one wishes to impose on
such a frame and these conditions are suggested by the problem under
consideration.  Here we shall choose the
conformal frame by demanding
that $T_{\alpha \beta}$ shall be consistent with the conservation property
of a stress-tensor.   We first consider a conformal transformation
\begin{equation}
g_{\alpha\beta} \rightarrow \bar{g}_{\alpha \beta} = \Omega^2 g_{\alpha \beta}~,
\label{3.9}\end{equation}
\begin{equation}
\psi(x) \rightarrow  \psi(x)~,
\end{equation}
where $\Omega$ is a nonvanishing, smooth spacetime function.  Then we
write Eq.(\ref{3.7}) in a conformal frame describing by the
metric $\bar{g}_{\alpha\beta}$ so that
\begin{equation}
\bar{\Sigma}_{\alpha\beta}=\bar{T}_{\alpha\beta}
+\frac{1}{48\pi} \bar{g}_{\alpha \beta} \Lambda~,
\label{3.10}
\end{equation}
where we have set
\begin{equation}
\bar{R}-\Lambda=0~,
\label{4.11}
\end{equation}
or, equivalently,
\begin{equation}
\nabla_{\gamma} \ln\Omega \nabla^{\gamma} \ln\Omega =\frac{1}{2}\Lambda\Omega^2
-\frac{1}{2}R+\frac{\Box \Omega}{\Omega}~.
\label{4.3}
\end{equation}
This is a differential equation which determines a particular conformal
factor characterizing a conformal frame in which the
relation (\ref{3.10}) holds.  Equation (\ref{4.11}) corresponds to a
two-dimensional analog of the vacuum Einstein field equations with a
cosmological constant\cite{man}.
This clearly emphasizes the significance of the gravitational coupling of
local states in the choice of a conformal frame as a physical one. We should also remark that
$\Lambda$ provides now the missing mass-scale mentioned previously in
connection with the Hadamard ansatz.  This mass-scale presents
itself through the root of an effective cosmological constant
in a two-dimensional gravitational theory.\\
To make a closer look at the relation (\ref{4.3}), let us investigate
it for a particular case in which the background geometry is described
by a two-dimensional Schwarzschild metric
\begin{equation}
ds^2=-(1-\frac{2GM}{r})dt^2 +(1-\frac{2GM}{r})^{-1} dr^2.
\label{sch}\end{equation}
In this case the trace
anomaly is $\Sigma^{\alpha}_{\alpha}=\frac{GM}{6\pi r^3}$.  To
solve Eq.(\ref{4.3}) in this particular case one needs some boundary
conditions.  One may intuitively expect that for a
sufficiently small cosmological constant the conformal frame, characterized
by (\ref{4.3}), and the asymptotic regions of the
background metric should have the same properties .  We therefore choose the
boundary conditions so that an asymptotic correspondence can be established
between the conformal frame and the background frame in the Schwarzschild
coordinate system.  Since the trace anomaly
vanishes when $r \rightarrow \infty$, Eq.(\ref{4.3})
reduces to
\begin{equation}
\nabla_{\gamma} \ln\Omega \nabla^{\gamma} \ln\Omega =\frac{1}{2}\Lambda\Omega^2
+\frac{\Box \Omega}{\Omega}~.
\label{4.4}
\end{equation}
For a sufficiently small $\Lambda$ and in a static case, in which $\Omega$ is
only a function of $r$, this equation has a simple solution,
namely $\Omega(r)=e^{(ar+b)}$ with $a$ and $b$ being arbitrary
constants. Applying this conformal factor to (\ref{sch}) gives in the
asymptotic regions
\begin{equation}
ds^2=e^{2(ar+b)} (-dt^2+dr^2)~.
\label{4.5}\end{equation}
The case $a=b=0$ corresponds
to the Minkowski spacetime and the conformal frame meets the background metric
for $r \rightarrow \infty$ in the Schwarzschild coordinate
system.  Thus $a=b=0$ characterizes the physically admissible conformal
factor.  When $a \neq 0$, a
coordinate transformation
$$
\eta=\frac{1}{a} e^{(ar+b)} \sinh(at)~,
$$
\begin{equation}
\xi=\frac{1}{a} e^{(ar+b)} \cosh(at)~,
\end{equation}
would change (\ref{4.5}) to
\begin{equation}
ds^2=-d\eta^2+d\xi^2~.
\label{4.6}\end{equation}
demonstrating that the coordinates $(t,r)$ are the Rindler-type
coordinates \cite{ri} associated with an accelerated observer in a flat
spacetime.  In this case the asymptotic correspondence between
the two frames can not be established in the Schwarzschild
coordinate system.

\section{Concluding remarks}
We have analyzed an analog of the Hadamard ansatz for the
specification of local
states associated with a linear quantum field conformally coupled to a
two-dimensional gravitational background.  For
studying the constraints emerging from this ansatz
we have employed a conformal invariant classical
scalar field.  The conformal invariance permits us to introduce
a conformal frame in which the
gravitational coupling of local states is established.  Such a frame
is constructed by relating the trace of the quantum stress-tensor to
a cosmological constant.  We have
shown that the gravitational coupling of the local states
leads to a two-dimensional gravitational theory.\\\\
{\bf Acknowledgment}\\
The authors would like to acknowledge the financial support of
the Office of Scientific Research of Shahid Beheshti University.\\\\  

\end{document}